\documentclass[conference]{IEEEtran}

\usepackage{hyperref}

\usepackage{forest}
\useforestlibrary{edges}
\usepackage{adjustbox}
\usepackage{subcaption}

\usepackage{algorithm}
\usepackage{listings}
\usepackage{caption}
\usepackage{amssymb}

\usepackage{tcolorbox}
\tcbuselibrary{listings,breakable}

\usepackage{balance}

\IEEEoverridecommandlockouts
\usepackage{cite}
\usepackage{amsmath,amssymb,amsfonts}
\usepackage{algorithmic}
\usepackage{graphicx}
\usepackage{textcomp}
\usepackage{xcolor}

\def\BibTeX{{\rm B\kern-.05em{\sc i\kern-.025em b}\kern-.08em
    T\kern-.1667em\lower.7ex\hbox{E}\kern-.125emX}}

\begin{document}

\title{TubeBEND: A Real-World Dataset for Geometry Prediction in Rotary Draw Bending}

\author{
\IEEEauthorblockN{Zeyneddin Oz\IEEEauthorrefmark{1}\textsuperscript{\textsection},
Jonas Knoche\IEEEauthorrefmark{2}\textsuperscript{\textsection},
Alireza Yazdani\IEEEauthorrefmark{1}\textsuperscript{\textsection},
Bernd Engel\IEEEauthorrefmark{2} and
Kristof Van Laerhoven\IEEEauthorrefmark{1}}
\IEEEauthorblockA{\IEEEauthorrefmark{1}Ubiquitous Computing, Department of Electrical Engineering and Computer Science, University of Siegen }
\IEEEauthorblockA{\IEEEauthorrefmark{2}Forming Technology Siegen, Department of Mechanical Engineering, University of Siegen \\
57076 Siegen, NRW, Germany \\
Email: {firstname.lastname}@uni-siegen.de}
}

\maketitle
\begingroup\renewcommand\thefootnote{\textsection}
\footnotetext{These authors contributed equally as first authors}
\endgroup

\begin{abstract}
This paper presents TubeBEND, a real-world dataset comprising 318 rotary tube bending processes, which were collected and sorted by experts from various fields to evaluate machine learning and signal analysis methods. The dataset addresses the industrial challenge of predicting the geometry of a first-stage bend, which can be beneficial for designing machine clamping molds for the second-stage bend in two-stage rotary draw bending. Some geometry criteria, such as the tube's final bent angle (or springback) and its cross-sectional deformation, are being recorded in this dataset. 
This dataset gives us the possibility to build and test machine learning models that can predict the geometry and help the machine operators with a better machine setup to optimize the tube's springback and deformation. Moreover, by recording some process parameters, such as tool movements and forces or torques applied to them, we deliver detailed information about their impacts on the final tube geometry. The focus of our work is to discover solutions that can replace traditional methods, such as trial-and-error or simulation-based predictions, by including experimental process variables in ML algorithms. 
Our dataset is publicly available at \href{https://github.com/zeyneddinoz/tubebend}{https://github.com/zeyneddinoz/tubebend} and \href{https://zenodo.org/records/16614082}{https://zenodo.org/records/16614082} as a benchmark to improve data-driven methods in this field.

\end{abstract}


\section{Introduction}

Bent tubes are important components in many fields, such as house HVAC (heating, ventilation, and air conditioning), aerospace, medical, automotive, and so on. They play an important role in supporting structures and transferring fluids. Achieving acceptable dimensional accuracy is challenging due to springback (the elastic recovery of bending angles after releasing machine tools) and other forming defects, such as collapse and wrinkling \cite{franz1988, lange1989}. These geometrical differences lead to errors in structural strength calculations, turbulent flow in fluid transformation, and assembly problems \cite{hoffmann2012handbuch, richtlinie2014rotation}.

Mathematical methods, physical trial-and-error bending processes, and Finite Element (FE) simulations have traditionally addressed these geometry mistakes \cite{lee2018stress, gan2004design, lingbeek2005development, karafillis1992tooling, li2013towards, borchmann2021regelung}. FE simulations can reduce physical tests, but they are not optimal because they cannot simulate all the possible forming conditions due to their simplified models, and they cannot fully capture the factors that occur in real production \cite{grossmann2009adjusting}. In addition, they require large computing times, which reduces the cost efficiency of Rotary Draw Bending (RDB) \cite{jonsson2020stamping, sun2011complex, yoshida2002elastic, groth2020methode}.

Machine Learning (ML)-based strategies offer an alternative solution; however, their accuracy, efficiency, and flexibility of their geometry control have not yet been studied \cite{ma2021machine, lu2022stretch, tao2016constitutive, zhang2022hierarchical, mayr2021data, sun2022digital}. Recent work on ML-based geometry prediction has used simulated data, achieving high accuracy and rapid prediction \cite{yazdani2025tube}. However, the effectiveness of these data-driven models for tube bending in real industrial applications is heavily dependent on the availability and quality of real-world data that capture the missing factors that deterministic simulations cannot fully replicate \cite{sun2022digital, kampker2018challenge, mayr2021data, zhang2022hierarchical, tao2016constitutive, lu2022stretch, ma2021machine}.

To fill this gap and improve the performance of prediction and compensation models, this paper presents a dataset derived from real-world experiments of industrial rotary tube bending processes, aiming to support advanced machine learning and signal analysis research in this area.

\section{Related Work}
The design and optimization of forming processes, particularly in complex operations like RDB, have significantly benefited from both advanced simulation techniques and, increasingly, the integration of real-world sensor data. This section reviews existing contributions, distinguishing between studies that primarily utilize simulated datasets and those based on real-world measurements, to highlight the current state of research and contextualize the TubeBEND dataset.

\textbf{Datasets from simulations:} Historically, the geometry of active tool surfaces in RDB has been designed using FE process simulations, often incorporating empirical knowledge \cite{miller2003tube}. However, these simulations frequently exhibit deviations between the calculated and actual component geometries due to unmodeled physical phenomena, parameter fluctuations, and transient effects within the forming system \cite{richtlinie2014rotation}. Such discrepancies are particularly pronounced in multi-bent components, where errors in a first bend can propagate and amplify in subsequent bends. For example, FE simulations of components with nested bends have revealed clear differences in geometry when Computer-Aided Design (CAD)-designed mold clamping dies are used compared to 'ideal' tapered mold clamping dies.

To address these deviations, numerous authors have developed methods for the simulation-based generation of working surface geometry to precompensate for component errors and dimensional inaccuracies such as springback \cite{grossmann2009adjusting, lee2018stress, gan2004design}. Two prominent approaches are the Displacement Adjustment (DA) method and the Spring Forward (SF) method. The DA method, initially published by \cite{gan2004design}, involves iteratively shifting tool mesh nodes in the FE environment according to springback displacement vectors until the desired component geometry is achieved. Subsequent developments have taken advantage of the NURBS functions to generate continuous three-dimensional tool surfaces \cite{lingbeek2005development}. In contrast, the SF method, introduced by \cite{karafillis1992tooling}, determines the contact forces in the closed tool state by FE simulation and then applies these forces in a linear-elastic simulation to derive the geometry of the tool. Both the DA and SF methods are only effective if the simulation models fully account for material-relevant influences, such as flow line curves, hardening behavior at work, and a decrease in modulus of elasticity with plastic deformation \cite{sun2011complex, yoshida2002elastic}. The researchers of \cite{grossmann2009adjusting} further emphasized the importance of including the parameters of the formation system, such as tool and machine elasticities, in the simulation models to achieve accurate and effective surface design. Specifically for RDB, the authors of \cite{li2013towards} presented a procedure for the automated generation of complete CAD datasets based on knowledge databases, parametric designs, and FE models.

Beyond tool design, simulated data have been a crucial source for training ML models, especially where real-world experimental data is scarce or costly \cite{tao2016constitutive, zhang2022hierarchical, mayr2021data, sun2022digital}. In \cite{yazdani2025tube}, for example, researchers used a Random Forest (RF) algorithm to calculate optimal forming paths for stretch bending. Another study used the same methodology with their entire training dataset derived from FE simulations that mapped input parameters to output quality \cite{lu2022stretch}. This enabled the trained RF model to replace computationally intensive FE simulations effectively. Similarly, \cite{zhang2022hierarchical} used hierarchical Gray Wolf Optimization with Support Vector Machines (GWO-SVM), trained on FE simulation data, to predict wrinkle formation during RDB, achieving an accuracy of 74.5\%. The authors of \cite{sun2022digital} also used digital twins based on FE models to perform real-time regressions and error classifications for the prediction of RDB springback, feeding measured data and machine parameters into a Multi-Task Learning (MTL) model.

\textbf{Datasets from actual processes:} Parallel to advances in simulation, there has been a significant trend towards integrating a large number of sensors into forming systems to capture physical influences from the real world on the process \cite{zorn2019potential, braunlich2002blecheinzugsregelung, hutter2021determination, wendler2021eddy, muhl2021soft, stebner2021system}. This sensor integration provides invaluable data for understanding and improving forming operations. Standard machine sensors, such as displacement or position sensors for axis control and force/torque or pressure sensors for load monitoring, have been widely used in processes like deep drawing to record process conditions \cite{zorn2019potential}. 

Specialized sensors have been developed to capture specific phenomena. For instance, sheet metal feed during deep drawing has been measured using eddy current, magnetic, or optical sensors installed in hold-down surfaces \cite{braunlich2002blecheinzugsregelung}. Local hold-down forces and temperatures have been measured using thin-film sensors applied to tool surfaces \cite{plogmeyer2020development}. In RDB, a line scanner was used to detect wrinkling in the area of smoother wrinkles \cite{borchmann2020situ}. Newer concepts involve soft sensors to record dynamic material properties by combining various sensor data (e.g., eddy currents and temperatures to calculate hardness and deformation degrees, or the magnetic Barkhausen effect) \cite{hutter2021determination, wendler2021eddy, muhl2021soft}. Direct hardness measurements are also used to calculate internal stresses during bending \cite{stebner2021system}. 

Real-world data are also increasingly utilized to train ML models. The \cite{mayr2021data} predicted geometric features based on the torque data of the machine axis drive motors during the production of hairpin stators. Their training dataset comprised 672 hairpins that were measured and linked to the torque and speed data of the four motors of the production machine, providing a clear example of a substantial real-world dataset. The author of \cite{sun2022digital} also leveraged measured data and machine parameters, available as time series, in an MTL model for RDB springback prediction. However, many previous research papers, despite their use of real-world data, explicitly noted limitations due to sparse real-time data acquisition, indicating that not all relevant process parameters could be comprehensively considered \cite{heftrich2018rotary, engel2013erweiterung, hinkel2013prozessfenster, borchmann2021regelung, hassan2017plasto}. This highlights a critical gap that the present project aims to address.

In the TubeBEND dataset, the aim was to employ an extensive real-world data collection strategy, including the recording of system variables in each bending cycle (e.g., axis travel and forces, tool movements, surface pressure, axial pull-out, component temperature, strain distribution, shape/position deviations). Additionally, semi-finished product characteristics (e.g., strength, surface quality, cross-sectional geometry) are measured per batch or at regular intervals. The objective is to establish an automated, time-synchronized measurement periphery. We were specifically tasked with creating a system to analyze the sensor data stream and preparing a reliable data pipeline for feature generation and common error classification. Thus, we created a "benchmark" reference dataset that provides a data-based representation of the process sequence. As a result, the TubeBEND dataset is particularly valuable, as it combines sensor readings with expert human knowledge and is currently publicly available at \href{https://github.com/zeyneddinoz/tubebend}{https://github.com/zeyneddinoz/tubebend} and \href{https://zenodo.org/records/16614082}{https://zenodo.org/records/16614082}.

\section{Dataset Collection}

\begin{figure*}[h]
    \centering
    \includegraphics[width=\linewidth]{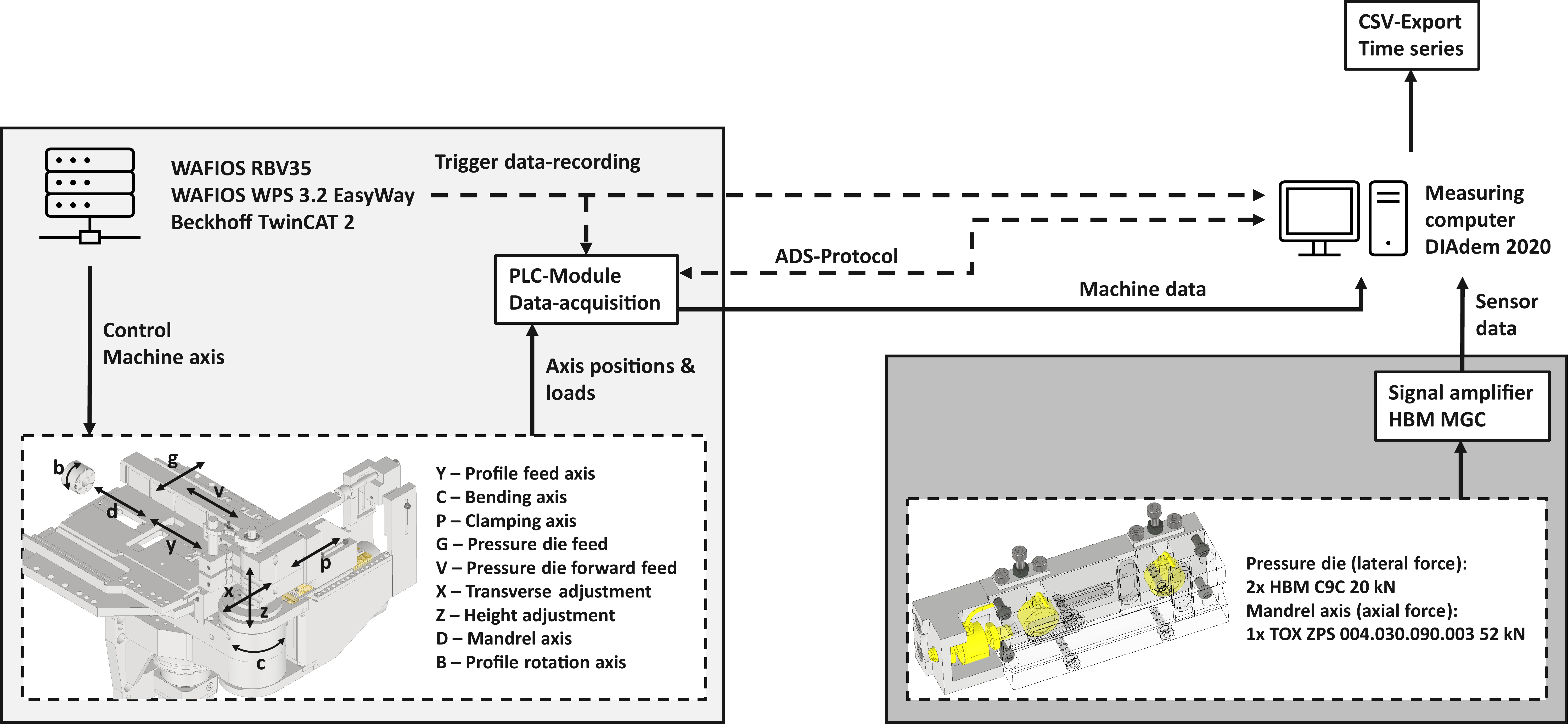}
    \caption{The data collection process delivers a variety of machine sensor data from integrated sensors in a 2016 WAFIOS RBV 35 tube bending machine, during the bending of a variety of stainless steel tubes.}  
  \label{fig:data-acquisition}
\end{figure*}

In this section, we summarize the data collection pipeline (overview in Fig.~\ref{fig:data-acquisition}) and the five subsections that follow: Bending Machine, Parameter Variations, Machine Data, Sensor Data, and Geometry. Briefly, these cover the RBV 35 hardware and control, the chosen process window (collet factors, pressure-die feed, mandrel timing), PLC-based axis/readout and recording cadence, dedicated force sensors with synchronized acquisition, and optical scanning plus geometry post-processing. The force and movement directions used in the dataset are defined in Fig.~\ref{fig:machine_directions}, and the geometric features (springback, collapse, section diameters, and the Linear-1 / Arc / Linear-2 cut regions) are illustrated in Fig.~\ref{fig:geo_data}.

\subsection{Bending Machine}
The RBV 35 tube bending machine, built in 2016 by WAFIOS AG, Reutlingen, is used for practical bending tests. The machine can be used for both rotary draw bending and three-roll push bending. The ten servo-controlled machine axes are programmed using the WAFIOS WPS 3.2 EasyWay control program, which is located on a separate industrial PC. The basis for the block structure of the control commands is DIN 66025 and the G and M commands contained therein. Individual commands are combined into lines in the program sequence. The control commands are converted into movements of the machine axis by the operating system. The Windows Control and Automation Technology (TwinCAT) 2 from Beckhoff Automation GmbH \& Co. KG, Verl, of a programmable logic controller (PLC). Tubes with an outer diameter of up to 35 mm, a maximum radius of 170 mm, and a maximum bending angle of 190° can be bent on up to three levels. The maximum bending moment is 6 kNm. By parameterizing the program sequence accordingly, selected process parameters such as bending angle, pressure die kinematics, mandrel retraction time, and collet kinematics can be quickly adjusted (see Fig.\ref{fig:data-acquisition}).

\subsection{Parameter Variations}

The production of a nested bend requires that the previous bend be flawless. For this reason, the parameter ranges examined are primarily determined by considering the failure characteristics in rotary draw bending as outlined in VDI 3430 \cite{vdi2014rotationszugbiegen}. In contrast to the formation of wrinkles on the inner arc, crack formation on the outer arc plays a minor role in the present bending task. Furthermore, the technical limitations of the bending machine must be considered.

The collet is a central element in controlling the flow of material. Depending on the bending task and the semi-finished product to be bent, the kinematics of the collet can be selected as pushing, pulling, or traveling. With pulling kinematics, the distance traveled by the collet during bending is shorter than the unwound length of the geometric center line of the tube in the bending area. This delay induces tensile stresses that counteract the formation of wrinkles on the inner bend \cite{borchmann2019sensitivity}. The ratio between the collet movement and the unwound geometric center line of the tube is referred to as the collet factor and is used to quantify the collet kinematics. The collet kinematics are limited downward by the holding capacity of the pneumatic collet within the collet. If the collet factor is below 0.85, there is a risk that the tube will slip through the collet. The upper limit of the parameter range for collet kinematics was set so that, with the tool setting used and a lateral pressure die feed of 0.9 mm, the onset of wrinkling is visible. This was the case with a collet factor of 0.95. This ensures that the limits of the process window are considered when evaluating the process parameters. The collet factors selected were 0.95, 0.92, 0.90, 0.87, and 0.85.

The pressure die acts as a counterbalance to the applied bending moment and prevents the tube from buckling. The pressure die lateral feed is achieved by placing a shim under the slide rail. The increments selected for the lateral feed of the pressure die were 0.9 mm, 0.6 mm, and 0.3 mm. As the study by \cite{yazdani2025tube} shows, a lateral feed of the pressure die in the range of 0 to 0.3 mm has no significant influence on the geometry during rotary draw bending \cite{yazdani2025tube}. For this reason, increments of 0.9 mm, 0.6 mm, as well as 0.3 mm were selected for the lateral feed of the pressure die. The pressure die kinematics can be selected in a manner comparable to the collet kinematics: stationary, traveling, as well as pushing. Pushing kinematics promotes the formation of wrinkles on the inner arc and is therefore not used.

\begin{figure}[t]
    \centering
    \includegraphics[width=\linewidth]{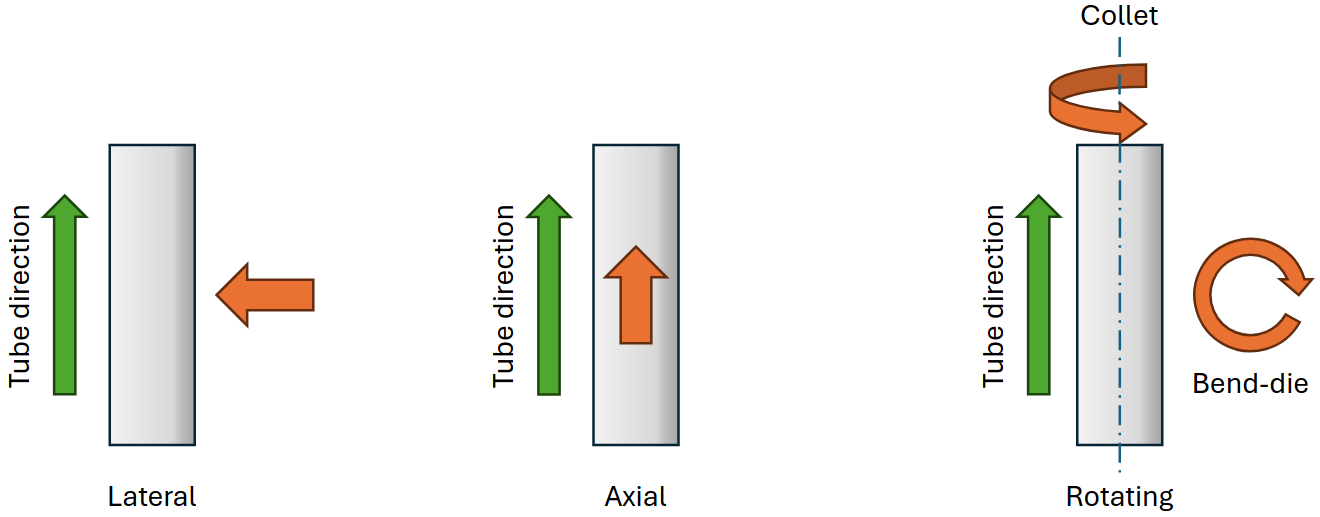}
    \caption{Directions of force and movements in relation to the tube axis as defined in the dataset for all machine components.}  
  \label{fig:machine_directions}
\end{figure}

\begin{figure*}[h]
    \centering
    \includegraphics[width=\linewidth]{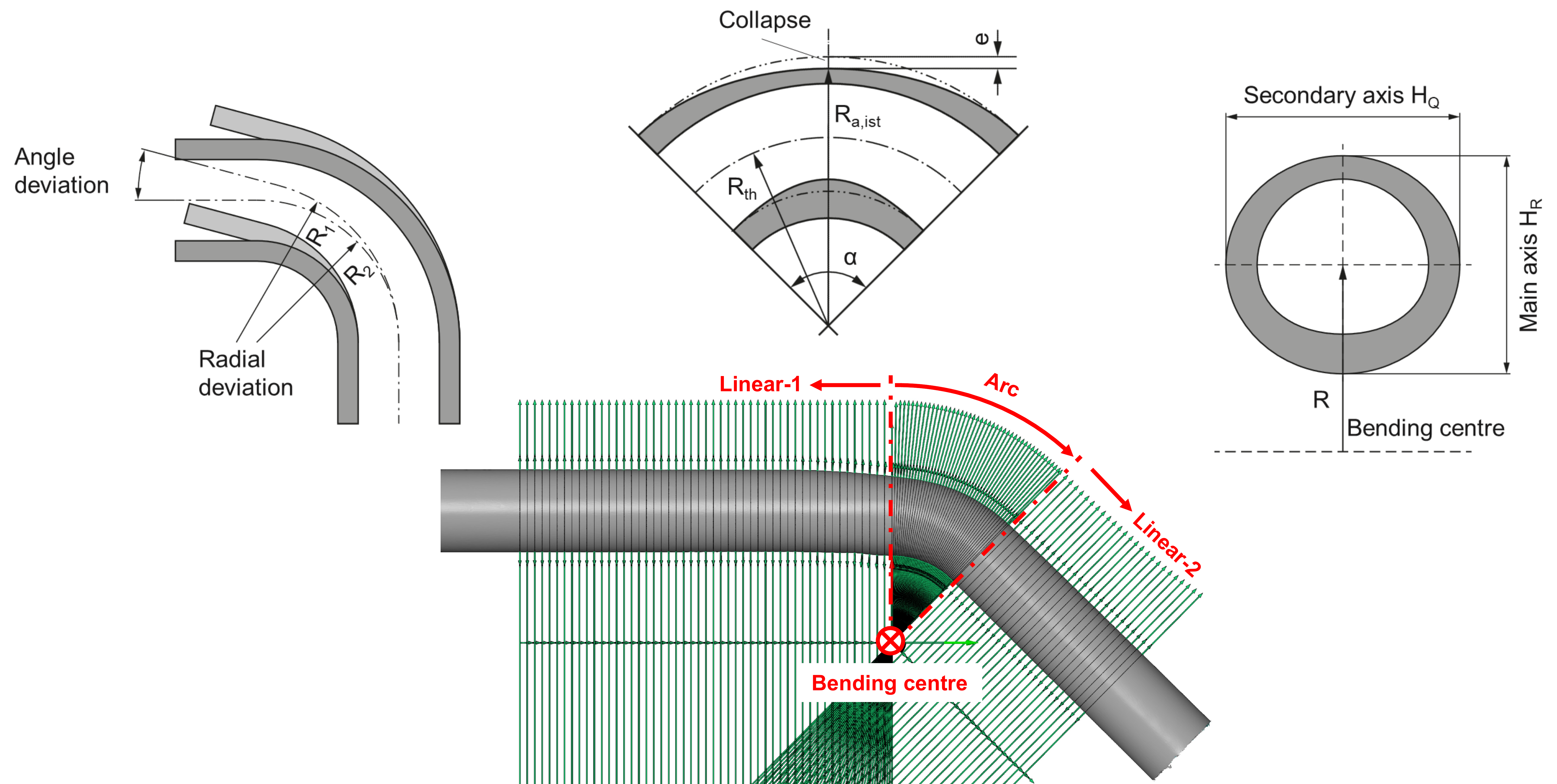}
    \caption{For each bending process, geometry data contains featues such as the springback (top left). For a series of cross-cuts along the tube (see the regions Linear-1, Arc, and Linear-2 in the bottom diagram), the collapse (middle left) and the diameters along the main and secondary bending axes (top right) are recorded.}  
  \label{fig:geo_data}
\end{figure*}

The mandrel serves as a support for the tube cross-section during bending. If the mandrel is pulled back too early,  more cross-section deformation can be expected. The timing of the mandrel retraction is specified in degrees before the end of the bend. The latest point for mandrel retraction is after completion of the bend, i.e., 0° before the end of the bend. Since the mandrel is also responsible for transferring the pressure die force to the wiper die, premature retraction can contribute to wrinkling on the inside arc. Furthermore, due to the opposite movement between the tube and the mandrel, increased stress due to friction is to be expected. Against this background and based on previous experience, the earliest point for mandrel retraction was set at 10° before the end of the bend. An intermediate point was defined as mandrel retraction at 5° before the end of the bend.

\subsection{Machine Data}

The machine data is read out during the bending process with the aid of a PLC module within the machine control system. The machine data to be read out includes the current positions and loads of the machine axes. The load on the machine axes is output based on the configuration of the PLC module in relation to the nominal or maximum torque of the axis motor. The Automation Device Specification (ADS) protocol is used for communication with the TwinCAT-based Beckhoff machine control system for the purpose of configuring data recording and reading the recorded machine data. The machine data is read in packets with a maximum size of 32767 values. The frequency at which the machine axes can be scanned and the data recorded is determined by two factors. On the one hand, it must be an integer multiple of the internal cycle time of the machine control, which is 2 ms. On the other hand, a packet released for transmission must be read out faster than is required for the next packet to be released. A release occurs when the defined size of the packet is reached, or the bending is completed, and the remaining data is to be read out. If a release for data transmission occurs while a previous packet is still being transmitted, it is partially overwritten, and the data is transmitted incorrectly. To ensure error-free reading of the machine data, a frequency of 20 Hz was specified for scanning the data recording. The start and end of the data recording can be specified via the G code of the bending program. Directions of force and movements, as defined in the dataset, are illustrated in Fig.~\ref{fig:machine_directions}.

\subsection{Sensor Data}

Since the load on the machine axes, especially those driven indirectly via gears, toothed belts, or spindles, only allows limited conclusions to be drawn about the tool forces acting on them, additional sensors were integrated to measure tool forces. A tension-compression sensor from TOX Pressotechnik SE \& Co. KG, Weingarten, was integrated into the mandrel bar to measure the mandrel force. The ZPS 004.030.090.003 sensor has a nominal force of 52 kN in the compression direction and a nominal force of 25 kN in the tension direction. In addition, two HBM C9B pressure transducers from Hottinger Brüel \& Kjaer GmbH, Darmstadt, with a nominal force of 20 kN each, were installed in the pressure die. All force sensors are connected to a computer for data recording via an MGC measuring amplifier from Hottinger Brüel \& Kjaer GmbH, Darmstadt. 

The computer has two PCIe-6531 measurement cards from National Instruments Corporation, Austin, which are connected to two BNC-2090A connection blocks from National Instruments Corporation, Austin. The measurement cards are configured, and the sensor data is recorded using DIAdem 2020 software from National Instruments Corporation, Austin. A frequency of 100 Hz was set for recording the sensor data. The start and end of data recording are analogous to the recording of machine data by commands within the G-code of the bending program. This ensures synchronized recording of machine and sensor data.

\subsection{Geometry}

The GOM ATOS Q 8M scanner, in conjunction with the MV500 measuring volume (500 x 370 x 320 mm³) from Carl Zeiss GOM Metrology GmbH, Braunschweig, is used to digitize the bent tubes. To capture the entire geometry of the bent tubes, a motorized turntable is used, which is integrated into the software of the ZEISS INSEPCT Pro Line 2023 scanner from Carl Zeiss GOM Metrology GmbH, Braunschweig. After confirmation of an initial scan, the bent tubes are digitized by automatic measurements during rotation of the turntable. The sensor works on the principle of fringe light projection. 

Before the digitized tubes can be evaluated, the scan data must be transformed from the sensor coordinate system into the machine coordinate system in accordance with VDI 3430. For this purpose, the bending plane is reconstructed from the center axes of the straight tube’s legs, and the intersection of the two axes is shifted along the bisector of the component angle to the center of the target bending radius. The target bending radius is 33 mm. The tube scans are then intersected along and perpendicular to the tube center line. The distances between the cuts are 1° in the curved area and 2 mm in the straight tube legs. The sections are used to determine the geometric features: collapse, secondary axis, and main axis.

The actual bending angle is calculated from the component angle enclosed by the center axes of the straight tube legs, minus 180°. The springback angle corresponds to the difference between the target and actual bending angles. The target bending angle is 47°. The geometry of the bent tubes is available in three degrees of abstraction. The lowest level of abstraction is an STL file based on the scan data of the respective tube. The discrete geometric features along the tube centerline, including the bending and springback angles, represent the highest level of abstraction. The coordinates of the sections along the tube centerline are available as a middle ground. The main geometric features considered in this study (springback, collapse, and cross-sectional diameters) are illustrated in Fig.~\ref{fig:geo_data}.

\section{Dataset Structure and Content}

This dataset is made from experimental data obtained directly from rotary tube bending production processes, all from the same tube material. It involves real-world data from 318 rotary tube bending processes, including 3 failed cases (IDs 1, 48, and 166). Its collection directly addresses the need for high-quality real-world data to train robust ML models. Unlike methods that use FE simulations or experiments in the laboratory, this dataset reflects the natural variability and complexity of real industrial production. This collection is approximately 800 MB in size and serves as a standard for many data-driven methods.

This dataset is arranged as a nested dictionary and serialized to a Python pickle file. Each leaf node is a Pandas dataframe with tabular measurements, and the keys (and subkeys) represent data types and categories. The structure resembles a tree, where internal nodes represent branches (data categories), and leaves represent tables.

Individual tables are selected by indexing the dictionary with the desired key path. For example, to extract the geometry key-characteristics table (the “Linear 1” side table) for experiment 47, we load the pickled nested-dictionary dataset and index the dictionary by the experiment identifier and the leaf key. Listing I shows the exact code used to load our dataset and obtain this table. This layout is easily extensible, allowing for straightforward navigation between different data types and experimental conditions.

\begin{tcblisting}{
    listing only,
    title=Listing I: An example for load data from the TubeBend dataset and accessing a specific table.,
    label=code:access_a_specific_table,
    listing options={
        language=Python,
        basicstyle=\scriptsize,
        breaklines=false,
        breakatwhitespace=false
    },
    left=2pt,
    top=1pt,          % Increased from 2pt to 6pt
    bottom=1pt,       % Added to balance the height
    boxsep=1pt,       % Increased from 2pt to 4pt
    arc=2pt,
    boxrule=0.5pt,
    colback=gray!10,
    colframe=black!80,
    fonttitle=\bfseries,
    halign=flush left,
    width=\linewidth  % Ensures the box uses the full line width
}
import pickle

# Load TubeBEND dataset as TB:
with open('experiments_process_and_results.pkl','rb') as f:
    TB = pickle.load(f)
    
# Show a specific data as a pandas dataframe:
TB['Exp_47']['geometry_data_key_characteristics_linear_1']

# Here, we selected the Linear 1 table under Key
# Characteristics for the Geometry Data of 47th Experiment.
\end{tcblisting}

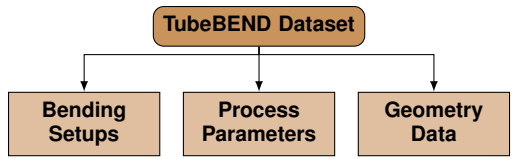
\begin{figure}[h]
\centering
\begin{forest}
for tree={
    draw,
    align=center,
    edge path={
        \noexpand\path[\forestoption{edge}]
        (!u.parent anchor) -- +(0,-3pt) -| (.child anchor)\forestoption{edge label};
    },
    l sep=6mm,
    s sep=3mm,
    anchor=center,
    parent anchor=south,
    child anchor=north,
    font=\sffamily\footnotesize,
    edge={->,>=latex},
    where level=0{
        fill=brown!85,  
        rounded corners,
        font=\sffamily\bfseries\footnotesize,
        inner sep=3pt,
        text width=2.6cm,
        minimum height=0.5cm
    }{},
    where level=1{
        fill=brown!50,  
        font=\sffamily\bfseries\footnotesize,
        inner sep=3pt,
        text width=1.8cm,
        minimum height=0.5cm
    }{},
    text centered
}
[TubeBEND Dataset  
    [Bending\\Setups]
    [Process\\Parameters]
    [Geometry\\Data]
]
\end{forest}
\caption{Main components of TubeBEND dataset.}
\label{fig:tube_bend_main_components}
\end{figure}

The dataset is comprehensively structured into three main categories (see Fig.~\ref{fig:tube_bend_main_components}), and is listed as follows:

\subsection{Bending Setups:} Details the initial configuration parameters serving as independent input variables for ML models, influencing deformation behavior (see Fig.~\ref{fig:bending_setups}):

\begin{itemize}
    \item \textbf{Tube:} Outer Diameter, and Wall Thickness.
    
    \item \textbf{Machine:} Bending Target Angle, Wiper Die Shortening, Pressure Die (Lateral Position, Distance, Boost), Mandrel (Position, Retraction Timing), Collet Boost, and Clamp Die Lateral Position.
\end{itemize}

\subsection{Process Parameters:} 
\begin{figure*}[!t]
\centering
\normalsize
\begin{adjustbox}{max width=\textwidth}
\begin{forest}
for tree={
    draw,
    align=center,
    edge path={
        \noexpand\path[\forestoption{edge}]
        (!u.parent anchor) -- +(0,-5pt) -| (.child anchor)\forestoption{edge label};
    },
    l sep=12mm,
    s sep=3mm,
    anchor=north,
    parent anchor=south,
    child anchor=north,
    font=\sffamily,
    if level=0{font=\bfseries\sffamily}{}, 
    edge={->,>=latex},
    where n children=0{fill=green!50}{} 
}
[Process Parameters, fill=brown!50, rounded corners
    [Loads, fill=brown!35
        [Machine, fill=brown!25
            [Bend Die, fill=green!25
                [Lateral]
                [Rotating]
                [Vertical]
            ]
            [Clamp Die Lateral]
            [Collet, fill=green!25
                [Axial]
                [Rotating]
            ]
            [Mandrel Axial]
            [Pressure Die, fill=green!25
                [Axial]
                [Lateral]
                [Left Axial]
            ]
        ]
        [Sensor, fill=brown!25
            [Mandrel Axial]
            [Pressure Die, fill=green!25
                [Lateral 1]
                [Lateral 2]
            ]
        ]
    ]
    [Movements, fill=brown!25
        [Bend Die, fill=green!25
            [Lateral]
            [Rotating]
            [Vertical]
        ]
        [Clamp Die Lateral]
        [Collet, fill=green!25
            [Axial]
            [Rotating]
        ]
        [Mandrel Axial]
        [Pressure Die, fill=green!25
            [Axial]
            [Lateral]
            [Left Axial]
        ]
    ]
]
\end{forest}
\end{adjustbox}
\caption{Tree diagram of the "Process Parameters" data structure. The brown root represents the overall parameter set; green boxes are columns grouped into tables. Under "Loads", there are two tables: "Machine" (10 columns) and "Sensor" (3 columns). The "Movements" table mirrors the "Machine" and contains 10 columns. Example column groups: Bend Die (Lateral, Rotating, Vertical), Collet (Axial, Rotating), Pressure Die (Axial, Lateral, Left Axial), and Mandrel Axial. In "Sensor", Pressure Die splits into "Lateral 1" and "Lateral 2".}
\label{fig:process_params}
\end{figure*}
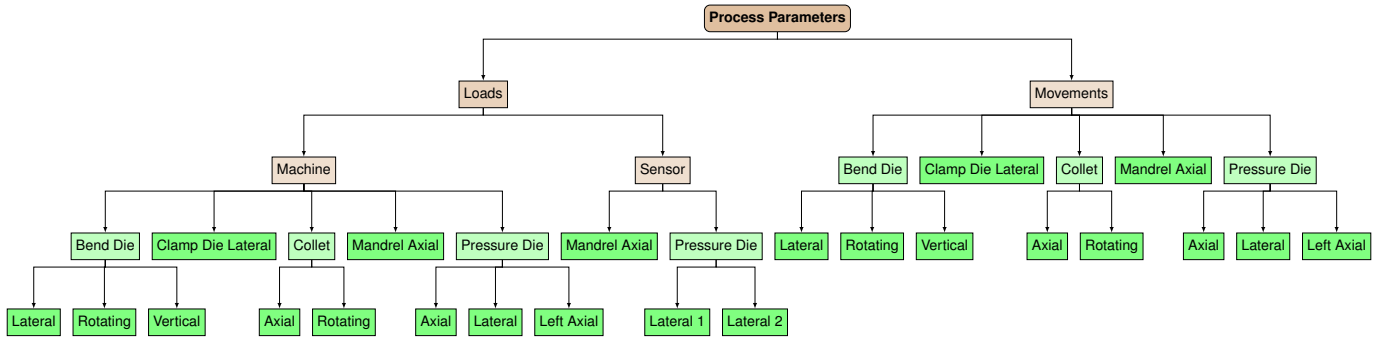

It captures dynamic measurements reflecting tool-tube interactions, which are critical for understanding process mechanics (see Fig.~\ref{fig:process_params}). Acquired from:

\begin{itemize}
    \item \textbf{Loads:}
    \begin{itemize}
        \item \textbf{Machine:} Bend Die (Lateral, Rotating, Vertical), Clamp Die Lateral, Collet (Axial, Rotating), Mandrel Axial, and Pressure Die (Axial, Lateral, Left Axial).
        \item \textbf{Sensor:} Mandrel Axial, and Pressure Die (Lateral 1, Lateral 2).
    \end{itemize}

    \item \textbf{Movements:} Bend Die (Lateral, Rotating, Vertical), Clamp Die Lateral, Collet (Axial, Rotating), Mandrel Axial, and Pressure Die (Axial, Lateral, Left Axial).
    
\end{itemize}

\subsection{Geometry Data:} 
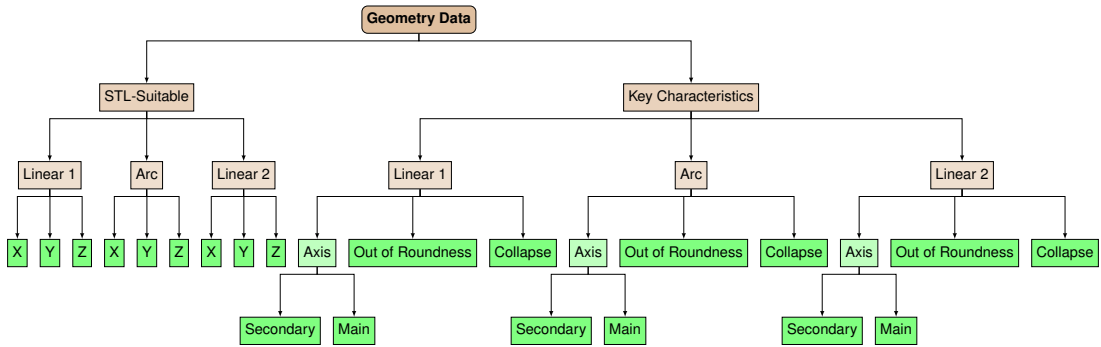
\begin{figure*}[!t]
\centering
\normalsize
\begin{adjustbox}{max width=0.8\textwidth}
\begin{forest}
for tree={
    draw,
    align=center,
    edge path={
        \noexpand\path[\forestoption{edge}]
        (!u.parent anchor) -- +(0,-5pt) -| (.child anchor)\forestoption{edge label};
    },
    l sep=12mm,
    s sep=3mm,
    anchor=north,
    parent anchor=south,
    child anchor=north,
    font=\sffamily,
    if level=0{font=\bfseries\sffamily}{},
    edge={->,>=latex},
    where n children=0{fill=green!50}{}  
}
[Geometry Data, fill=brown!50, rounded corners  
    [STL-Suitable, fill=brown!35  
        [Linear 1, fill=brown!25  
            [X]
            [Y]
            [Z]
        ]
        [Arc, fill=brown!25       
            [X]
            [Y]
            [Z]
        ]
        [Linear 2, fill=brown!25  
            [X]
            [Y]
            [Z]
        ]
    ]
    [Key Characteristics, fill=brown!35  
        [Linear 1, fill=brown!25  
            [Axis, fill=green!25  
                [Secondary]
                [Main]
            ]
            [Out of Roundness]
            [Collapse]
        ]
        [Arc, fill=brown!25  
            [Axis, fill=green!25  
                [Secondary]
                [Main]
            ]
            [Out of Roundness]
            [Collapse]
        ]
        [Linear 2, fill=brown!25  
            [Axis, fill=green!25  
                [Secondary]
                [Main]
            ]
            [Out of Roundness]
            [Collapse]
        ]
    ]
]
\end{forest}
\end{adjustbox}
\caption{Tree diagram of the "Geometry Data" structure. The brown root groups geometry into two categories: "STL-Suitable" (three tables: Linear 1, Arc, Linear 2), which hold 3D coordinate values (X, Y, Z), and "Key Characteristics" (matching Linear 1, Arc, Linear 2), which list measured properties. Each Key Characteristics table contains three main categories: "Axis" (with sub-categories Secondary and Main), "Out of Roundness", and "Collapse".}
\label{fig:geometry_data}
\end{figure*}

This provides detailed geometric information of the bent tube (see Fig.~\ref{fig:geometry_data}), and can be listed as follows:

\begin{itemize}

    \item \textbf{STL-Suitable:} Mesh data for 3D reconstruction of three sections:
    \begin{itemize}
        \item \textbf{Linear 1:} A series of data from one endpoint of the arc.
    
        \item \textbf{Arc:} Datasets ranging from 0 to the bending target angle, increasing every 1 degree.
    
        \item \textbf{Linear 2:} A series of data from another endpoint of the arc.
    \end{itemize}
        
    \item \textbf{Key Characteristics:} Quantitative features for Linear 1, Arc, Linear 2 sections:
    \begin{itemize}

        \item \textbf{Linear 1:} Secondary Axis, Main Axis, and Out of Roundness.
    
        \item \textbf{Arc:} Secondary Axis, Main Axis, and Out of Roundness.
    
        \item \textbf{Linear 2:} Secondary Axis, Main Axis, and Out of Roundness.

    \end{itemize}
        
\end{itemize}

\begin{figure*}[!t]
\centering
\normalsize
\begin{adjustbox}{max width=0.8\textwidth}
\begin{forest}
for tree={
    draw,
    align=center,
    edge path={
        \noexpand\path[\forestoption{edge}]
        (!u.parent anchor) -- +(0,-5pt) -| (.child anchor)\forestoption{edge label};
    },
    l sep=12mm,
    s sep=3mm,
    anchor=north,
    parent anchor=south,
    child anchor=north,
    font=\sffamily,
    if level=0{font=\bfseries\sffamily}{},
    edge={->,>=latex},
    where level=0{fill=brown!50, rounded corners}{},  
    where level=1{fill=green!5}{},                  
    where level=2{fill=green!25}{},                  
    where n children=0{fill=green!50}{}              
}
[Bending Setups
    [Tube
        [Outer Diameter]
        [Wall Thickness]
    ]
    [Machine
        [Bending Target Angle]
        [Wiper Die Shortening]
        [Pressure Die
            [Lateral Position]
            [Distance]
            [Boost]
        ]
        [Mandrel
            [Position]
            [Retraction Timing]
        ]
        [Collet Boost]
        [Clamp Die Lateral Position]
    ]
]
\end{forest}
\end{adjustbox}
\caption{Tree diagram of the "Bending Setups" table structure. The brown root box indicates the table; green boxes show hierarchical column groups. "Outer Diameter" and "Wall Thickness" are columns under the Tube group, while the Machine group contains "Bending Target Angle", "Wiper Die Shortening", "Collet Boost", "Clamp Die Lateral Position", and the subgroups Pressure Die and Mandrel. Pressure Die expands into three columns ("Lateral Position", "Distance", "Boost"); Mandrel expands into two ("Position", "Retraction Timing").}
\label{fig:bending_setups}
\end{figure*}
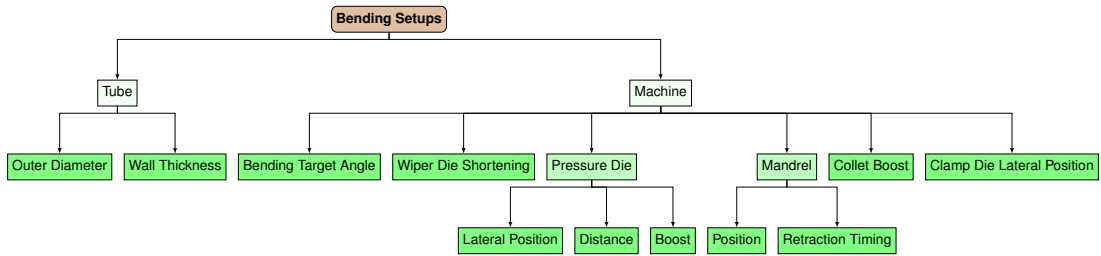

In summary, the following keys in the dataset dictionary file give related tables:

\begin{itemize}
    \item[{\scriptsize $\rightarrow$}] {\small \texttt{bending\_setups}}
    \item[{\scriptsize $\rightarrow$}] {\small \texttt{process\_parameters\_loads\_machine}}
    \item[{\scriptsize $\rightarrow$}] {\small \texttt{process\_parameters\_loads\_sensor}}
    \item[{\scriptsize $\rightarrow$}] {\small \texttt{process\_parameters\_movements}}
    \item[{\scriptsize $\rightarrow$}] {\small \texttt{geometry\_data\_stl\_suitable\_linear\_1}}
    \item[{\scriptsize $\rightarrow$}] {\small \texttt{geometry\_data\_stl\_suitable\_arc}}
    \item[{\scriptsize $\rightarrow$}] {\small \texttt{geometry\_data\_stl\_suitable\_linear\_2}}
    \item[{\scriptsize $\rightarrow$}] {\small \texttt{geometry\_data\_key\_characteristics\_linear\_1}}
    \item[{\scriptsize $\rightarrow$}] {\small \texttt{geometry\_data\_key\_characteristics\_arc}}
    \item[{\scriptsize $\rightarrow$}] {\small \texttt{geometry\_data\_key\_characteristics\_linear\_2}}
\end{itemize}

\section{Dataset Usage and Applications}

This real-world dataset offers potential for various applications in advanced manufacturing and data science:

\textbf{1- Machine Learning and Signal Analysis Benchmarking:} This TubeBEND dataset is explicitly designed to serve as a benchmark for the development and evaluation of ML algorithms, such as Random Forest Regression or Artificial Neural Networks (e.g., enhanced Particle Swarm Optimization-Backpropagation (PSOBP) neural networks and Long Short-Term Memory (LSTM), and signal analysis methods for industrial processes. Its "real-world" aspect, which comes from industrial production, captures the impacts that are essential for creating workable and reliable solutions.

\textbf{2- Springback Prediction and Compensation:} By including target bending angles and the resulting final bent angle in the geometry dataset, it is possible to train machine learning models for prediction, create compensation models, and create a data-driven "bridge" connecting input variables to the resulting springback.

\textbf{3- Cross-sectional Deformation Analysis:} Predicting and controlling cross-sectional deformation is essential for product quality and is made possible by recording precise geometry-specific measures, which include Secondary Axis, Main Axis, Out of Roundness, and Collapse.

\textbf{4- Physics-Informed Neural Networks (PINNs):} The combination of high-frequency process data (forces, movements) and high-quality geometric outcomes (springback, collapse) makes this dataset ideal for training PINNs. These models embed physical laws (e.g., deformation, elasticity, plasticity) into the learning process as soft constraints or loss terms. This helps address data sparsity, improves generalization beyond the observed parameter range, and ensures that the predictions remain physically consistent. PINNs can be used to capture stress–strain relationships or to build accurate surrogate models for springback and cross-sectional deformation.

\textbf{5- Process Parameter Optimization:} By utilizing the wide variety of bending configurations, researchers can find the best settings that reduce defects and improve product quality. The most predictive characteristics, including collet boost and mandrel retraction timing, can be found by using the importance analysis of features, which helps guide the choice of the best tactics \cite{yazdani2025tube}.

\textbf{6- Anomaly Detection and Process Monitoring:} The high-frequency sensor and machine data provide a perfect benchmark for developing real-time anomaly detection algorithms. Models can be trained to identify signatures of impending failures (e.g., severe wrinkling, tool overload, or machine faults) from the force and position signals before they result in a scrap part, enabling predictive interventions.

\textbf{7- Root Cause Analysis for Defects:} By correlating specific parameter sets (e.g., low collet factor + early mandrel retraction) with geometric results (high collapse, wrinkles), the TubeBEND dataset can be used to build classifiers and models that not only predict defects but also identify the most likely contributing process parameters, helping to rapid diagnosis and correction on the shop floor.

\textbf{8- Optimizing Mold Design:} The TubeBEND dataset can be used to build models to predict the geometry of the tube, which can later help to improve clamp mold designs, potentially lowering manual rework, material damage, and repetitive compensation in production.

\textbf{9- Predictive Maintenance:} The recorded loads on the machine axes (e.g., bending arm, pressure die) reflect the mechanical stress on the system. This data can be used to model wear and tear, helping to predict maintenance needs for critical components such as servomotors, gears, and ball screws before they fail, reducing unplanned downtime.

\textbf{10- Multi-Objective Optimization:} The TubeBEND dataset allows for exploration of trade-offs between competing quality goals. For instance, a parameter set that minimizes collapse might increase springback. Algorithms can be developed to find Pareto-optimal solutions that best balance multiple objectives such as geometric accuracy, tool wear, and cycle time.

\textbf{11- Handling Stochastic Impacts:} The TubeBEND dataset directly reflects the impacts that are inherent in industrial production and are challenging to represent deterministically, in contrast to datasets that are only produced from simulations. This makes it possible to create prediction models that are more reliable and applicable to real-world production settings.

\textbf{12- Enabling Closed-Loop Control and Digital Twins:} The information can serve as a basis for incorporating machine learning models into production processes, looking for self-adaptive improvement, and investigating closed-loop control techniques for defect control and springback compensation. In addition, it supports the development of digital twins by offering real-world process data for visualization, model coupling, and defect classification \cite{sun2022digital}.

\section{Discussion and Outlook}

This real-world dataset provides a resource for advancing data-driven approaches in rotary tube bending. It provides a chance to create and evaluate machine learning and signal analysis techniques by including bending configurations, dynamic process parameters, and the resulting geometric features from 318 real industrial steel tube bending processes. The TubeBEND dataset can be used to improve production adaptations, optimize tooling designs, regulate cross-sectional deformation, and improve springback compensation.

Since the efficiency of machine learning models depends on the amount and types of data they receive, increasing their dataset is important to reduce error changes and develop strong compensation strategies. This means that the current dataset, although large, may still limit the generalizability of the model for all industrial scenarios.

To address this problem, future work can combine FE simulation-based experiments with real-world production data to create a larger "hybrid big-data source" and determine the optimal dataset size that balances prediction accuracy with computational speed. Ultimately, this expanded base can facilitate the development of closed-loop control systems through self-adaptive model enhancement utilizing real-time production data. This progress from the traditional "trial and error" method could accelerate intelligent, adaptive tube-bending production, reduce production costs, and minimize compensation errors.

\section*{Acknowledgment}

This work was funded by the Deutsche Forschungsgemeinschaft (DFG, German Research Foundation) – Project-IDs 506589320 and 520256321.

\balance

\begin{thebibliography}{10}
\providecommand{\url}[1]{#1}
\csname url@samestyle\endcsname
\providecommand{\newblock}{\relax}
\providecommand{\bibinfo}[2]{#2}
\providecommand{\BIBentrySTDinterwordspacing}{\spaceskip=0pt\relax}
\providecommand{\BIBentryALTinterwordstretchfactor}{4}
\providecommand{\BIBentryALTinterwordspacing}{\spaceskip=\fontdimen2\font plus
\BIBentryALTinterwordstretchfactor\fontdimen3\font minus \fontdimen4\font\relax}
\providecommand{\BIBforeignlanguage}[2]{{%
\expandafter\ifx\csname l@#1\endcsname\relax
\typeout{** WARNING: IEEEtran.bst: No hyphenation pattern has been}%
\typeout{** loaded for the language `#1'. Using the pattern for}%
\typeout{** the default language instead.}%
\else
\language=\csname l@#1\endcsname
\fi
#2}}
\providecommand{\BIBdecl}{\relax}
\BIBdecl

\bibitem{franz1988}
W.-D. Franz, \emph{Maschinelles Rohrbiegen -- Verfahren und Maschinen}.\hskip 1em plus 0.5em minus 0.4em\relax D{\"u}sseldorf: VDI-Verlag, ISBN 3-18-400814-2, 1988.

\bibitem{lange1989}
K.~Lange, \emph{Umformtechnik -- Handbuch f{\"u}r Industrie und Wissenschaft, Bd. 3: Blechbearbeitung}.\hskip 1em plus 0.5em minus 0.4em\relax Stuttgart: Springer, ISBN 978-3-662-10686-0, 1989.

\bibitem{hoffmann2012handbuch}
H.~Hoffmann, R.~Neugebauer, and G.~Spur, \emph{Handbuch umformen}.\hskip 1em plus 0.5em minus 0.4em\relax Hanser M{\"u}nchen, 2012, vol.~2.

\bibitem{richtlinie2014rotation}
VDI-Richtlinie, ``Rotationszugbiegen von profilen,'' \emph{Verband Deutscher Ingenieure}, 2014.

\bibitem{lee2018stress}
S.~Lee, M.~Eder, D.~Maier, and W.~Volk, ``Stress-based compensation of geometrical deviations in metal forming,'' in \emph{Congress of the German Academic Association for Production Technology}.\hskip 1em plus 0.5em minus 0.4em\relax Springer, 2018, pp. 647--656.

\bibitem{gan2004design}
W.~Gan and R.~Wagoner, ``Die design method for sheet springback,'' \emph{International Journal of Mechanical Sciences}, vol.~46, no.~7, pp. 1097--1113, 2004.

\bibitem{lingbeek2005development}
R.~Lingbeek, J.~Huetink, S.~Ohnimus, M.~Petzoldt, and J.~Weiher, ``The development of a finite elements based springback compensation tool for sheet metal products,'' \emph{Journal of Materials Processing Technology}, vol. 169, no.~1, pp. 115--125, 2005.

\bibitem{karafillis1992tooling}
A.~Karafillis and M.~Boyce, ``Tooling design in sheet metal forming using springback calculations,'' \emph{International journal of mechanical sciences}, vol.~34, no.~2, pp. 113--131, 1992.

\bibitem{li2013towards}
H.~Li, H.~Yang, and K.~Liu, ``Towards an integrated robust and loop tooling design for tube bending,'' \emph{The International Journal of Advanced Manufacturing Technology}, vol.~65, no.~9, pp. 1303--1318, 2013.

\bibitem{borchmann2021regelung}
L.~Borchmann, ``Regelung des werkstoffflusses zur erhoehung der bauteilqualitat beim rotationszugbiegen,'' Ph.D. dissertation, Dissertation, Siegen, Universit{\"a}t Siegen, 2021, 2021.

\bibitem{grossmann2009adjusting}
K.~Gro{\ss}mann, H.~Wiemer, A.~Hardtmann, L.~Penter, and S.~Kriechenbauer, ``Adjusting the contact surface of forming tools in order to compensate for elastic deformations during the process,'' in \emph{7th European LS-DYNA Conference}, 2009.

\bibitem{jonsson2020stamping}
C.-J. Jonsson, R.~Stolt, and F.~Elgh, ``Stamping tools for sheet metal forming: Current state and future research directions,'' \emph{Transdisciplinary Engineering for Complex Socio-Technical Systems--Real-Life Applications}, pp. 281--290, 2020.

\bibitem{sun2011complex}
L.~Sun and R.~Wagoner, ``Complex unloading behavior: Nature of the deformation and its consistent constitutive representation,'' \emph{International Journal of Plasticity}, vol.~27, no.~7, pp. 1126--1144, 2011.

\bibitem{yoshida2002elastic}
F.~Yoshida, T.~Uemori, and K.~Fujiwara, ``Elastic--plastic behavior of steel sheets under in-plane cyclic tension--compression at large strain,'' \emph{International journal of plasticity}, vol.~18, no. 5-6, pp. 633--659, 2002.

\bibitem{groth2020methode}
S.~Groth, \emph{Methode zur Produktplanung beim Freiformbiegen}.\hskip 1em plus 0.5em minus 0.4em\relax Shaker Verlag, 2020.

\bibitem{ma2021machine}
J.~Ma, H.~Li, G.~Chen, T.~Welo, and G.~Li, ``Machine learning (ml)-based prediction and compensation of springback for tube bending,'' in \emph{Forming the future: proceedings of the 13th international conference on the technology of plasticity}.\hskip 1em plus 0.5em minus 0.4em\relax Springer, 2021, pp. 167--178.

\bibitem{lu2022stretch}
K.~Lu, T.~Zou, J.~Luo, D.~Li, and Y.~Peng, ``Stretch bending process design by machine learning,'' \emph{The International Journal of Advanced Manufacturing Technology}, vol. 120, no.~1, pp. 781--799, 2022.

\bibitem{tao2016constitutive}
Z.-J. Tao, H.~Yang, H.~Li, J.~Ma, and P.-F. Gao, ``Constitutive modeling of compression behavior of tc4 tube based on modified arrhenius and artificial neural network models,'' \emph{Rare Metals}, vol.~35, no.~2, pp. 162--171, 2016.

\bibitem{zhang2022hierarchical}
S.~Zhang, Y.~Yuan, Z.~Wang, Y.~Lin, L.~Jiang, and M.~Fu, ``A hierarchical prediction method based on hybrid-kernel gwo-svm for metal tube bending wrinkling detection,'' \emph{The International Journal of Advanced Manufacturing Technology}, vol. 121, no.~7, pp. 5329--5342, 2022.

\bibitem{mayr2021data}
A.~Mayr, P.~R{\"o}ll, D.~Winkle, M.~Enzmann, B.~Bickel, and J.~Franke, ``Data-driven quality monitoring of bending processes in hairpin stator production using machine learning techniques,'' \emph{Procedia CIRP}, vol. 103, pp. 256--261, 2021.

\bibitem{sun2022digital}
C.~Sun, Z.~Wang, S.~Zhang, T.~Zhou, J.~Li, and J.~Tan, ``Digital-twin-enhanced metal tube bending forming real-time prediction method based on multi-source-input mtl,'' \emph{Structural and Multidisciplinary Optimization}, vol.~65, no.~10, p. 296, 2022.

\bibitem{yazdani2025tube}
A.~Yazdani, J.~Knoche, B.~Engel, and K.~Van~Laerhoven, ``Tube geometry prediction in rotary draw bending process using random forest regression,'' \emph{at-Automatisierungstechnik}, vol.~73, no.~4, pp. 223--231, 2025.

\bibitem{kampker2018challenge}
A.~Kampker, K.~D. Kreisk{\"o}ther, M.~K. B{\"u}ning, and P.~Treichel, ``Challenge of hairpin technology technology boost for oems and plant manufacturers,'' \emph{ATZelektronik worldwide}, vol.~13, no.~5, pp. 54--59, 2018.

\bibitem{miller2003tube}
G.~G. Miller, \emph{Tube forming processes: a comprehensive guide}.\hskip 1em plus 0.5em minus 0.4em\relax Society of Manufacturing Engineers, 2003.

\bibitem{zorn2019potential}
W.~Zorn, L.~Hamm, R.~Elsner, and W.-G. Drossel, ``Potential of the force distribution measurement in deep drawing processes for increasing the process quality,'' \emph{Int. J. Mech. Eng. Robot. Res}, vol.~8, no.~3, pp. 449--453, 2019.

\bibitem{braunlich2002blecheinzugsregelung}
H.~Br{\"a}unlich, ``Blecheinzugsregelung beim tiefziehen mit niederhalter-ein beitrag zur erh{\"o}hung der prozessstabilit{\"a}t,'' Ph.D. dissertation, Verlag Wissenschaftliche Scripten, 2002.

\bibitem{hutter2021determination}
S.~H{\"u}tter, R.~Lafarge, J.~Simonin, G.~Mook, A.~Brosius, and T.~Halle, ``Determination of microstructure changes by eddy-current methods for cold and warm forming applications,'' \emph{Advances in Industrial and Manufacturing Engineering}, vol.~2, p. 100042, 2021.

\bibitem{wendler2021eddy}
F.~Wendler, R.~Munjal, M.~Waqas, R.~Laue, S.~H{\"a}rtel, B.~Awiszus, and O.~Kanoun, ``Eddy current sensor system for tilting independent in-process measurement of magnetic anisotropy,'' \emph{Sensors}, vol.~21, no.~8, p. 2652, 2021.

\bibitem{muhl2021soft}
F.~M{\"u}hl, M.~Knoll, M.~Khabou, S.~Dietrich, P.~Groche, and V.~Schulze, ``Soft sensor approach based on magnetic barkhausen noise by means of the forming process punch-hole-rolling,'' \emph{Advances in Industrial and Manufacturing Engineering}, vol.~2, p. 100039, 2021.

\bibitem{stebner2021system}
S.~C. Stebner, D.~Maier, A.~Ismail, S.~Balyan, M.~D{\"o}lz, B.~Lohmann, W.~Volk, and S.~M{\"u}nstermann, ``A system identification and implementation of a soft sensor for freeform bending,'' \emph{Materials}, vol.~14, no.~16, p. 4549, 2021.

\bibitem{plogmeyer2020development}
M.~Plogmeyer, G.~Gonz{\'a}lez, V.~Schulze, and G.~Br{\"a}uer, ``Development of thin-film based sensors for temperature and tool wear monitoring during machining,'' \emph{tm-Technisches Messen}, vol.~87, no.~12, pp. 768--776, 2020.

\bibitem{borchmann2020situ}
L.~Borchmann, P.~Frohn-S{\"o}rensen, and B.~Engel, ``In situ detection and control of wrinkle formation during rotary draw bending,'' \emph{Procedia Manufacturing}, vol.~50, pp. 589--596, 2020.

\bibitem{heftrich2018rotary}
C.~Heftrich, R.~Steinheimer, and B.~Engel, ``Rotary-draw-bending using tools with reduced geometries,'' \emph{Procedia Manufacturing}, vol.~15, pp. 804--811, 2018.

\bibitem{engel2013erweiterung}
B.~Engel and C.~Mathes, \emph{Erweiterung der Prozessf{\"a}higkeit des Rotationszugbiegens durch ein alternatives Faltengl{\"a}tterkonzept: Ergebnisse eines Vorhabens der industriellen Gemeinschaftsforschung...}\hskip 1em plus 0.5em minus 0.4em\relax Europ{\"a}ische Forschungsges. f{\"u}r Blechverarbeitung (EFB), 2013.

\bibitem{hinkel2013prozessfenster}
M.~Hinkel, \emph{Prozessfenster f{\"u}r das Spannen von Rohrprofilen beim Rotationszugbiegen unter Ber{\"u}cksichtigung der Tribologie}.\hskip 1em plus 0.5em minus 0.4em\relax Shaker, 2013.

\bibitem{hassan2017plasto}
H.~R. Hassan, \emph{Plasto-mechanical Model of Tube Bending in Rotary Draw Bending Process}.\hskip 1em plus 0.5em minus 0.4em\relax Shaker Verlag, 2017.

\bibitem{vdi2014rotationszugbiegen}
V.~3430, ``Rotationszugbiegen von profilen,'' \emph{.}, 2014.

\bibitem{borchmann2019sensitivity}
L.~Borchmann, C.~Kuhnhen, P.~Frohn, and B.~Engel, ``Sensitivity analysis of the rotary draw bending process as a database of digital equipping support,'' \emph{Procedia Manufacturing}, vol.~29, pp. 592--599, 2019.

\end{thebibliography}
\bibliographystyle{IEEEtran}

\end{document}